\documentstyle[aps,preprint,epsfig,tighten]{revtex}
\newcommand{\BE}{\begin{equation}}
\newcommand{\EE}{\end{equation}}
\newcommand{\BEA}{\begin{eqnarray}}
\newcommand{\EEA}{\end{eqnarray}}
\begin{document}
\draft
\title{Effects of Disoriented Chiral Condensates on Two- and
Three-Pion Correlations of Relativistic Nuclear Collisions }
\author{Hiroki Nakamura$^{(1)}$ and Ryoichi Seki$^{(2,3)}$}
\address{
${}^{(1)}$ Department of Physics,
Waseda University, Tokyo 169-8555, Japan \\
${}^{(2)}$ Department of Physics, California State University,
Northridge, CA 91330 \\
${}^{(3)}$ W.~K.~Kellogg Radiation Laboratory, 106-38, 
California Institute of Technology, \\
Pasadena, CA 91125} 
\date{\today}
\maketitle
\begin{abstract}
Two- and three-pion correlations are investigated 
in cases when disoriented chiral condensate (DCC) occurs.
A chaoticity and weight factor are used
as measures of two- and three-pion correlations, and  
the various models for DCC are investigated.
Some models are found to yield the chaoticity and weight factor 
in a reasonable agreement with recent experimental data.

\end{abstract}

\bigbreak

\section{Introduction}
It is expected that 
chiral symmetry will be restored within the high temperature region
generated by high-energy nuclear or proton collisions.
The phase transition associated with the restoration may be
of a nonequilibrium nature, caused by the rapid expansion and cooling
of the high-temperature region.  Through such a phase transition, 
the chiral order-parameter could have a direction in the isospin space,
different from that in the true vacuum.  This phenomenon is called
disoriented chiral condensate (DCC) and has been intensely investigated
in the last several years\cite{rajagopal}.
The most prominent signature of DCC is expected to be that the fractional
neutral-pion number obeys a specific probability distribution, 
a simple version of which is
\begin{equation}
\label{eq:dist}
P(f)=\frac{1}{2\sqrt{f}},
\end{equation}
derived by assuming that the direction of condensates is isotropic
in the isospin space\cite{rajagopal,Bjorken}.  Here, 
 $f$ is the fractional neutral-pion number, a ratio of
the neutral-pion number and the total pion number, given by 
$f = \frac{N_{\pi_0}}{N_{\pi_+}+N_{\pi_-}+N_{\pi_0}}$.
The distribution predicts a large probability
that the number of the neutral pions is observed
much less than that of the charged pions.  Apparently such events,
known as Centauro events\cite{centauro}, are observed
in cosmic-ray experiments. 
Accelerator experiments, such as the Minimax and WA98 experiments, 
have attempted to measure this distribution\cite{minimax,wa98}, but no clear
signature of such a distribution has been observed.

In relativistic heavy-ion collisions, two-pion correlations have been used 
to obtain information about the size and shape of the pion-emitting source,
based on the Hanbury-Brown Twiss (HBT) effect.
Three-pion correlations are expected
to provide possibly new information about the source\cite{hz,ns}.
The first experimental data regarding
the three-pion correlations has been recently
reported\cite{na44}.  In our previous paper\cite{ns2}, we investigated
effects of a not completely chaotic source on the two- and
three-pion correlations, without referring to any specific dynamics
involved in the phase transition.  If DCC should occur, a
natural question would be how well effects of DCC could be observed in pion
interferometry. In this paper, we address that question.

It has been pointed out that the two-pion correlations yield anomalous 
values when the pions are emitted from a single coherent source exhibiting
DCC\cite{ggm,hm,gavin}.  
Perhaps it is reasonable to assume the pion-emitting source associated
with DCC to be coherent.  
As we will discuss in Sec. III, the pions emitted from such a source 
do not exhibit the HBT effect and do not serve as a means 
of identifying DCC.  It would be naive, however, to assume 
that the pions would be emitted by a single coherent source exhibiting DCC
and by nothing else.  For example, there could be more than one coherent
source associated with DCC, or coherent sources could appear 
with the background of a chaotic source.  It is quite reasonable
to expect that such a domain structure might occur at the appearance of 
DCC\cite{Bjorken,gavin,rajagopal}.
We investigate the effects on both the two- and three-pion correlations,
of various possible characteristics of a source 
having domain structure.

A note on terminology.  As mentioned above, we treat a source or
multiple sources of DCC as coherent.
If a source is not associated with DCC, we will call it generic.
A generic source or generic sources can be either coherent or 
chaotic.  A generic source that is combined with DCC sources will, 
however, be regarded as chaotic in this work.
In our previous work\cite{ns2}, we investigated models in which 
coherent and chaotic generic sources are combined in various ways.
The DCC models in Sec. IV have the same structure
as the models in the previous work, except that a DCC source or DCC sources  
have been replaced by a coherent source or coherent sources, respectively.  
In figures, we will compare numerical results obtained by the use of 
the two groups of models.  When we do so, we will introduce abbreviated 
names of the models for clear identification of curves in the figures.

In Sec. \ref{sec2},  pion spectra and correlation
functions are reviewed, and a chaoticity and weight factor
are defined as measures of the strength of two- and three-pion
correlations.
We investigate correlations for a single DCC domain in Sec. \ref{sec3},
and for a more realistic pion source involving DCC in Sec. \ref{sec4}.
In Sec.\ref{sec5},
we compare our results with a recent experiment
and present a summary of this work.

\section{Pion spectra and correlation functions}
\label{sec2}

In this section, for clarity, we define the pion spectra and correlation
functions that will be used in the rest of this work.  Our definitions
are conventional.
 
One-, two- and three-pion spectra are given by
\begin{mathletters}
\label{eqs:spect}
\begin{eqnarray}
W_1^i(p) &=& \langle{a^i_p}^\dagger{a^i_p}\rangle,\\
W_2^{ij}(p_1,p_2) &=&  \langle{a^i_{p_1}}^\dagger{a^j_{p_2}}^\dagger
            {a^i_{p_1}}{a^j_{p_2}}\rangle, \\
W_3^{ijk}(p_1,p_2,p_3) &=&  \langle{a^i_{p_1}}^\dagger
            {a^j_{p_2}}^\dagger
            {a^k_{p_3}}^\dagger{a^i_{p_1}}
            {a^j_{p_2}}{a^k_{p_3}}\rangle,
\end{eqnarray}
\end{mathletters}
where $i$, $j$, and $k$ stand for $+$, $-$, or $0$, specifying the pion charge
such as $\pi_+$, $\pi_-$, or $\pi_0$, respectively.
All momenta are on-shell, $p^0 = \sqrt{{\mathbf{p}}^2+m^2}$, above and
hereafter.
$\langle ...\rangle$ is formally $\langle\psi|...|\psi\rangle$ in terms of 
the quantum state, $|\psi\rangle$, or 
${\mathrm{Tr}}\{\hat{\rho}...\}$ in terms of the density matrix, $\hat{\rho}$.
In this work we introduce various models in place of the formal
definition.

In terms of the spectra, we define correlation functions as
\begin{mathletters}
\label{eqs:cdef}
\begin{eqnarray}
C_2^{ij}(p_1,p_2) &=& \frac{W_2^{ij}(p_1,p_2)}{W_1^i(p_1)W_1^j(p_2)},\\
C_3^{ijk}(p_1,p_2,p_3) &=& \frac{W_3^{ijk}(p_1,p_2,p_3)}
    {W_1^i(p_1)W_1^j(p_2)W_1^k(p_3)}.
\end{eqnarray}
\end{mathletters}
As in \cite{gkw}, in order to exclude the effect of multiplicity fluctuation,
we use normalized correlation functions for identical pions:
\begin{mathletters}
\BEA
\label{eq:nc2def}
\overline{C}_2^{ii}(p_1,p_2)
    &=& \frac{\langle \hat{n}_i \rangle^2}
     {\langle \hat{n}_i(\hat{n}_i-1) \rangle}
       C_2^{ii}(p_1,p_2), \\
\overline{C}_3^{iii}(p_1,p_2,p_3) &=&
  \frac{\langle \hat{n}_i \rangle^3}
   {\langle \hat{n}_i(\hat{n}_i-1)(\hat{n}_i-2) \rangle}
     C_3^{iii}(p_1,p_2,p_3),
\EEA
\end{mathletters}
where
\begin{mathletters}
\BEA
\langle \hat{n}_i \rangle &=& \int d^3p W_1^i(p), \\
\langle \hat{n}_i(\hat{n}_i-1) \rangle 
&=& \int d^3p_1d^3p_2 W_2^{ii}(p_1,p_2), \\
\langle \hat{n}_i(\hat{n}_i-1)(\hat{n}_i-2) \rangle &=& 
\int d^3p_1d^3p_2d^3p_3 W_3^{iii}(p_1,p_2,p_3).
\EEA
\end{mathletters}
Here, $\hat{n}_i$ is the number operator, 
$\hat{n}_i=\int d^3p a_p^{i\dagger}a_p^i$.

We define chaoticity, $\lambda_{ij}(p)$, and weight factor, $\omega_{ijk}(p)$,
as
\begin{eqnarray}
\label{eq:chaoticity}
\lambda_{ij}(p) &=& \left(
  \overline{C}_2^{ij}(p_1,p_2) - 1\right)_{p_1=p_2=p},\\
\omega_{ijk}(p) &=& \left.\frac{\overline{C}_3^{ijk}(p_1,p_2,p_3)-1
   - (\overline{C}_2^{ij}(p_1,p_2)
   +\overline{C}_2^{jk}(p_2,p_3) 
   +\overline{C}_2^{ki}(p_3,p_1) - 3)
  }
  {2\sqrt{ 
    (\overline{C}_2^{ij}(p_1,p_2) - 1)
    (\overline{C}_2^{jk}(p_2,p_3) - 1)
    (\overline{C}_2^{ki}(p_3,p_1) - 1)
  }}\right|_{p_1=p_2=p_3=p}. \nonumber \\
\end{eqnarray}
$\lambda_{ij}(p)$ and $\omega_{ijk}(p)$ are measures of the strength of
the two- and three-pion correlations.  Note that $\lambda_{ij}$ and 
$\omega_{ijk}$ are defined for zero relative momenta.
$\omega_{ijk}$ has been extracted from the recent experiment for small 
relative momenta\cite{na44}, but the difference is immaterial since it 
is a slowly varying function of the relative momenta\cite{hz,ns}.
When $\lambda_{ij}(p)$ is unity or zero, we call the pion-emitting source
chaotic or coherent, respectively.  For a chaotic source, $\omega_{ijk}(p)$
is unity.

\section{a single coherent source associated with DCC}
\label{sec3}

In this section, we examine the case of pions that are emitted from a single
coherent source associated with DCC.  The correlation functions in this case
can be derived in two different ways, by the isosinglet method and
by the generating functional method. 
The isosinglet method has been previously used 
to obtain the anomalous values of the two-pion correlation functions\cite{kt}.
The generating functional method\cite{Bjorken} is a more systematic approach,
and we will apply it to more complicated cases in Sec. IV.  Though
the two methods yield the same result in this simple case, we will 
show both for the sake of comparison.
Both methods show that no HBT effect occurs 
in this case.

\subsection{The Isosinglet Method}
\label{iso-singlet}

An isosinglet state vector of $2N$ pions is given by
\begin{equation}
  |\psi\rangle = \frac{1}{\sqrt{(2N+1)!}} {I^\dagger}^N|0\rangle,
\end{equation}
where
\begin{eqnarray}
 I 
   &=& 2 a_+ a_- -a_0^2 .
\end{eqnarray}
$I$ is an isosinglet operator\cite{hornsilver,kt,ggm}.
Since the creation and annihilation operators are of a single mode, 
the momentum variables are not explicitly shown here.
The probability of finding $2n$ neutral pions is
\begin{eqnarray}
P_{2n,2N} &=& \frac{(N!)^2 2^{2N} (2n)!}{(2N+1)! (n!)^2 2^{2n}} \nonumber\\
&\to& \frac{1}{2\sqrt{n/N}} \quad (N,n \gg 1).
\end{eqnarray}
This large $N$ and $n$ limit is consistent with Eq. (\ref{eq:dist}).
The positive-pion spectra come out to be 
\begin{mathletters}
\begin{eqnarray}
W_1^+ &= &\frac{2}{3}N,\\
W_2^{++} &=& \frac{8}{15}N ( N - 1 ), \\
W_3^{+++} &=& \frac{16}{35}N ( N - 1) ( N - 2 ).
\end{eqnarray}
\end{mathletters}
In the large $N$ limit, the  two- and three-pion correlations
for various combinations of pion charges defined in Eqs. (\ref{eqs:cdef})
are then 
\begin{mathletters}
\label{eqs:cors}
\begin{eqnarray}
\label{eq:cha1}
&&C_2^{\pm\pm} =
C_2^{\pm\mp} = \frac{6}{5} ,\\
&&C_2^{\pm 0} = \frac{4}{5},\\
\label{eq:cha1end}
&&C_2^{00} = \frac{9}{5},\\
&&C_3^{\pm\pm\pm} =
C_3^{\pm\pm\mp} =\frac{54}{35}, \\
&&C_3^{\pm\pm 0} =
C_3^{\pm\mp 0} = \frac{54}{105}, \\
&&C_3^{\pm 00} =  \frac{27}{35}, \\
&&C_3^{000} = \frac{27}{7},
\label{eq:wf1}
\end{eqnarray}
\end{mathletters}
The anomalous values of the two-pion correlations,
Eqs. (\ref{eq:cha1})--(\ref{eq:cha1end}), agree with those given 
previously\cite{ggm}.  We see that the three-pion correlation functions 
also yield anomalous values.  

DCC gives correlation functions different from those of a coherent 
generic source.  We note, however, that the normalized correlation 
functions, $\overline{C}$'s, are the same as those of a coherent generic 
source, all being unity.  
The reason for this is seen as follows: The isosinglet method is applied 
for a single mode (momentum), and yields the same $W$'s for all modes.  
The normalization condition, Eq. (5), thus guarantees that all 
$\overline{C}$'s as defined in Eq. (4) are unity, 
the same value as those of the coherent generic source.
The pions emitted from a single coherent DCC source do not then exhibit 
the HBT effect and do not serve as a means of identifying DCC.

\subsection{The Generating Functional Method}
\label{sec:pdcc}

In order to achieve the isosinglet property of DCC,
pion spectra should be averaged over all directions of isospin 
space\cite{Bjorken}.
We evaluate correlation functions with $c$-number source-current formalism.

In $c$-number source-current formalism, a generating functional
method is useful\cite{apw}. We define the generating functional
(slightly differently from {}\cite{apw}) as
\BE
G[z_i^*(p),z_i(p)]= \left\langle 
 \exp\left\{\sum_{i}\int \frac{d^3{\mathrm{p}}}{\sqrt{(2\pi)^3\cdot 2p^0}}
\left(z_i^*(p)J_i^*(p)+z_i(p)J_i(p)\right)
  \right\} \right\rangle_J,
\label{eq:gf}
\EE
where $J_i(p)$ is the $c$-number source current for the pion charge, 
$i = +, -,$ or 0.  $\langle ... \rangle_J$
is the statistical average of the fluctuation of $J_i(p)$.
Using this generating functional, we can obtain
\BE
\langle a_{p_1}^{i_1\dagger}
a_{p_2}^{i_2\dagger} \cdots a_{p_n}^{i_n\dagger}
 a_{q_1}^{j_1}
a_{q_2}^{j_2} \cdots a_{q_m}^{j_m}
\rangle
= 
\left.
\frac{\delta^{n+m} G[z_i^*(p),z_i(p)]}
{\delta z^*_{i_1}(p_1)\cdots \delta z^*_{i_n}(p_n)
\delta z_{j_1}(q_1)\cdots \delta z_{j_m}(q_m)}
\right|_{z_i(p)=0}.
\EE

Following \cite{Bjorken}, we set the source current of
each kind of pion, $J_i(p)$,  as
\begin{eqnarray}
J_0(p) &=& J(p)n_3 \\
J_\pm(p) &=& J(p) \frac{n_1\mp i n_2}{\sqrt{2}},
\end{eqnarray}
where $\mathbf n $ is a unit vector in isospin space. 
The ratio of neutral pions, $f$, is given by
\begin{equation}
f = n_3^2,
\end{equation}
and $|n_\pm|^2 = (1-f)/2$.
$n_i$ is independent of momentum because
we assume that all modes of momenta condense in the same direction in
isospin space.
Due to the isotropic distribution of the unit vector,
 the average that appears in Eq. (\ref{eq:gf})
becomes
\BE
\langle\cdots\rangle_J = \int \frac{d^3{\mathbf{n}}}{4\pi}
\delta(|{\mathbf{n}}|-1) \cdots.
\EE
When we treat identical pions, the integrand of the right-hand 
side depends on only $n^2_3$ or $|n_\pm|^2$. In such cases,
we can replace the integral as
\BE
 \int \frac{d^3{\mathbf{n}}}{4\pi}\delta(|{\mathbf{n}}|-1)\cdots = 
\int_0^1 df
P(f) \cdots,
\EE
where $P(f)$ is defined in Eq. (\ref{eq:dist}).

The neutral-pion spectrum for DCC is obtained as
\begin{eqnarray}
W_1^0(p) &=& \int_0^1P(f)df   n_3^2 W(p) \nonumber \\
&=& \frac{1}{3}  W(p),
\end{eqnarray}
where
\begin{equation}
W(p) = \frac{1}{2p^0 (2\pi)^3}|J(p)|^2. 
\end{equation}
For the charged pions, we obtain
\begin{equation}
W_1^\pm(p)=W_1^0(p).
\end{equation}
Similarly, we can derive two-pion spectra,
\begin{mathletters}
\begin{eqnarray}
W_2^{\pm\pm}(p_1,p_2) &=&
W_2^{\pm\mp}(p_1,p_2) = \frac{2}{15} W(p_1)W(p_2),\\
W_2^{0\pm}(p_1,p_2) &=& \frac{1}{15} W(p_1)W(p_2),\\
W_2^{00}(p_1,p_2) &=& \frac{1}{5} W(p_1)W(p_2).
\end{eqnarray}
\end{mathletters}
Three-pion spectra  are 
\begin{mathletters}
\begin{eqnarray}
W_3^{\pm\pm\pm}(p_1,p_2,p_3)& =& 
W_3^{\pm\pm\mp}(p_1,p_2,p_3) = 
\frac{2}{35} W(p_1)W(p_2)W(p_3),\\
W_3^{\pm\pm 0}(p_1,p_2,p_3) &=&
W_3^{\pm\mp 0}(p_1,p_2,p_3) =
 \frac{2}{105}  W(p_1)W(p_2)W(p_3),\\
W_3^{\pm 00}(p_1,p_2,p_3) &=& \frac{1}{35}  W(p_1)W(p_2)W(p_3),\\
W_3^{000}(p_1,p_2,p_3) &=& \frac{1}{7}   W(p_1)W(p_2)W(p_3).
\end{eqnarray}
\end{mathletters}
Equations (23) and (24) yield the same correlation functions as 
those obtained by the use of the isosinglet method, Eqs. (\ref{eqs:cors}).
Furthermore, $W(p)$ of Eq. (21) corresponds to the pion-number density 
for the mode (momentum) $p$, 
and $W_n$'s are $n$-products of $W(p)$'s.  
Accordingly, when the normalization condition of Eq. (5) is imposed, 
the $\overline{C}$'s automatically become unity, which is the same value 
of $\overline{C}$'s as in the coherent generic case.  We thus see that 
though the correlation functions are momentum-dependent in this method 
while they are independent in the isosinglet method, 
both methods yield the same $C$'s and also $\overline{C}'{\rm s} = 1$.    

When the pions are emitted from more than one coherent source with DCC, 
or they are also emitted from a chaotic generic source, 
the HBT effect appears and 
DCC could be identified.  We will discuss these cases in the following section.

\section{A source with DCC domain}
\label{sec4}

In relativistic heavy-ion collisions, 
it would not be realistic to consider all the pions being emitted from 
one large DCC source, as in the model in the previous section.
In this section, we investigate three models that we expect to be 
more realistic. 
In the first model, the source consists of one DCC source and one chaotic 
generic source, while in the second model the source consists 
of multiple DCC domains.
The third model is a combination of the previous two models, 
multiple DCC domains with one chaotic generic source. 
The HBT effect appears in these models and 
helps identifying the DCC signature in the pion interferometry.

\subsection{A Partially Coherent DCC Source (PC-DCC)}
\label{sec:pcdcc}

In this subsection,
we introduce a partially coherent source with DCC\cite{Bjorken,gavin}.
Source current, $J_i(p)$, is separated into two parts as
\begin{equation}
J_i(p) = J_D(p) n_i + J^i_g(p),
\end{equation}
where $J_D(p)$ is the DCC source current and $J^i_g(p)$ is the generic one.
We assume that the DCC source is coherent and 
that the generic source is chaotic.
The average in Eq. (\ref{eq:gf}) becomes 
\BE
\langle\cdots\rangle_J=
\int \frac{d^3{\mathbf{n}}}{4\pi}\delta(|{\mathbf{n}}|-1)\int
\cdots
\prod_{i=1}^3
{\cal P}_i[J_g^{i*}(p),J_g^i(p)]
{\cal D}J_g^i(p)
{\cal D}J_g^{i*}(p),
\EE
where ${\cal P}_i[J_g^{i*}(p),J_g^i(p)]$ is a distribution functional
of $J_g^i(p)$ and assumed to have a Gaussian form, as in Ref.\cite{apw},
so that the higher-order moment of $J_g^i(p)$ is represented
 by the second-order moment, for example,
\BEA
\lefteqn{\langle J_g^{i*}(p_1)J_g^{i*}(p_2)J_g^i(p_1)J_g^i(p_2) \rangle_J
=} \nonumber \\
&& \langle J_g^{i*}(p_1)J_g^i(p_1) \rangle_J
 \langle J_g^{i*}(p_2)J_g^i(p_2) \rangle_J
+ \langle J_g^{i*}(p_1)J_g^i(p_2) \rangle_J
 \langle J_g^{i*}(p_2)J_g^i(p_1) \rangle_J.
\EEA

At first, we concentrate on the case of positive pions.
We introduce here several real functions, $f_{ij}$, 
$\phi_{ij}$, $F^+_{ij}$ and $\Phi^+_{ij}$.
These functions are defined as follows\cite{hz},
\begin{eqnarray}
 f_{ij}\exp(i\phi_{ij}) &\equiv&
   \frac{1}{(2\pi)^3\sqrt{2{p^0_1}\cdot 2{p^0_2}}}
    J_D^*(p_i)J_D(p_j) \\ 
 F^+_{ij}\exp(i\Phi^+_{ij}) &\equiv& 
   \frac{1}{(2\pi)^3\sqrt{2{p^0_1}\cdot 2{p^0_2}}} 
   \langle J_g^{+*}(p_i)J^+_g(p_j) \rangle_{J^+_g}, 
\label{eq:fij}
\end{eqnarray} 
where
\BE
\langle\cdots \rangle_{J^+_g} =\int \cdots {\cal P}_+[J_g^{+*}(p),J_g^+(p)]
{\cal D}J_g^+(p)
{\cal D}J_g^{+*}(p).
\EE
Hereafter, for simplicity, we denote $F^+_{ij}$ as $F_{ij}$.
We obtain the positive-pion spectrum, 
\begin{eqnarray} 
 W_1^+(p_1) &=& \int_0^1dfP(f) W_f^+(p_1) \nonumber \\
&=& \frac{1}{3} f_{11}+F_{11},
\end{eqnarray} 
where
\begin{equation}
W^i_f(p_1) = |n_i|^2f_{11}+F^i_{11}.
\end{equation}
The two-pion spectrum is given by
\begin{eqnarray}
W_2^+(p_1,p_2) &=&\int_0^1P(f)df
 \left\{ W_f^+(p_1)W_f^+(p_2)
 + R_f^{+}(p_2,p_3)
   \right\}\nonumber \\
&=& \frac{1}{45}f_{11}f_{22}
 + W_1^+(p_1)W_1^+(p_2)
 +R^+(p_1,p_2),
\label{eq:twodcc}
\end{eqnarray}
where 
\begin{eqnarray}
R_f^{i}(p_1,p_2) =  {F^i_{12}}^2+ 2 |n_i|^2
   f_{12}F^i_{12}\cos\left(\Phi_{12}-\phi_{12}\right) \\
R^{i}(p_1,p_2) =  {F^i_{12}}^2+ \frac{2}{3}
   f_{12}F^i_{12}\cos\left(\Phi_{12}-\phi_{12}\right).
\end{eqnarray}
The first term in the right-hand side of  Eq. (\ref{eq:twodcc})
does not appear in the case of the partially coherent generic source\cite{ns2},
signifying the new aspect in the partially coherent DCC case.
The two-pion correlation function becomes
\begin{equation}
C_2^{++}(p_1,p_2) 
= 1+\frac{1}{5}(1-\epsilon(p_1))(1-\epsilon(p_2))
+\frac{R^+(p_1,p_2)}{W_1^+(p_1)W_1^+(p_2)}
\end{equation}
where $\epsilon(p)$ is the ratio of the generic pion number and 
the total pion number:
\begin{equation}
\epsilon(p_1) = \frac{F_{11}}{\frac{1}{3}f_{11}+F_{11}}
\label{eq:ratio}
\end{equation}
In this case, the normalization of Eq. (\ref{eq:nc2def}) is not 
unity in the limit of the infinite relative momenta.
This is because the term $\frac{1}{5}(1-\epsilon(p_1))(1-\epsilon(p_2))$
is momentum-dependent.
In the following, we proceed by neglecting that the momentum-dependence of
$\epsilon(p)$ is independent of $p$, so that the correlation function is
normalized as in Eq. (\ref{eq:nc2def}):
\BE
\overline{C}_2^{++}(p_1,p_2) = 1+\frac{1}{1+\frac{1}{5}(1-\epsilon)^2}
   \frac{R^+(p_1,p_2)}{W_1^+(p_1)W_1^+(p_2)}.
\EE
The chaoticity for normalized correlation function, Eq. (\ref{eq:chaoticity}), 
is
\begin{equation}
{\lambda}_{++} = \frac{\epsilon(2-\epsilon)}
   {1+\frac{1}{5}(1-\epsilon)^2}.
\end{equation}
This chaoticity is different from that of a partially coherent generic source:
\BE
\lambda_{generic} = \epsilon(2-\epsilon),
\EE
which is evaluated in \cite{hz}.
In Fig. \ref{fig:chap}, we show the chaoticity as a function of the ratio,
$\epsilon$, for a partially coherent DCC source (PC-DCC).  
For comparison, we also show in the figure the chaoticity for a partially 
coherent generic source (PC-G) that has been previously 
investigated\cite{ns2,hz,gkw}.

The three-pion spectrum is evaluated as
\begin{eqnarray}
W_3^{+++}(p_1,p_2,p_3) &=& \int_0^1dfP(f)\bigg\{
  W_f^+(p_1) W_f^+(p_2) W_f^+(p_3) 
 + \sum_{(a,b,c)}W_f^+(p_a)R_f^{+}(p_b,p_c)
 \nonumber \\
& +& 2\bigg[F_{12}F_{23}F_{31} \cos(\Phi_{12}+\Phi_{23}+\Phi_{31})
\nonumber \\
&&
    +\frac{1-f}{2}\sum_{(a,b,c)}
      f_{ab}F_{bc}F_{ca}\cos(\phi_{ab}+\Phi_{bc}+\Phi_{ca})
   \bigg]
  \bigg\}
\nonumber
\\
 &=&
  W_1^+(p_1)W_1^+(p_2)W_1^+(p_3)
 + \sum_{(a,b,c)}W_1^+(p_a)R^{+}(p_b,p_c)
\nonumber \\ &&
 + 2\bigg[F_{12}F_{23}F_{31} \cos(\Phi_{12}+\Phi_{23}+\Phi_{31})
\nonumber \\ &&
    +\frac{1}{3}\sum_{(a,b,c)}
      f_{ab}F_{bc}F_{ca}\cos(\phi_{ab}+\Phi_{bc}+\Phi_{ca})
   \bigg]
\nonumber \\
&+&\frac{19}{945}f_{11}f_{22}f_{33}
+\frac{1}{45}\sum_{(a,b,c)}
   f_{aa}f_{bb}F_{cc}
+\frac{2}{45}\sum_{(a,b,c)}
    f_{aa}f_{bc}F_{bc}\cos(\Phi_{bc}-\phi_{bc}).
\end{eqnarray}
The weight factor for normalized three-pion correlation functions
is
\begin{eqnarray}
\omega_{+++} &=& 
\left[
\frac{3\epsilon(2-\epsilon)+2\epsilon^2(3-2\epsilon)
  +\frac{6}{5}\epsilon(1-\epsilon)^2}
{1+\frac{19}{35}(1-\epsilon)^3+\frac{3}{5}(1-\epsilon)^2\epsilon}
 - 3{\lambda}_{++}
\right]\frac{{ \lambda}_{++}^{-\frac{3}{2}}}{2}
\end{eqnarray}
Figure \ref{fig:wfp} illustrates the weight factors for
a partially coherent DCC source and a partially coherent generic source
as functions of $\epsilon$.
The weight factor of a partially coherent DCC model becomes 
negative for small $\epsilon$.
This is the aspect of the DCC case that characteristically differs 
from the generic
case:
\BE
\omega_{generic} = \sqrt{\epsilon}\frac{3-2\epsilon}{(2-\epsilon)^{3/2}},
\EE
which is always positive.

The chaoticity and weight factor for normalized neutral-pion correlations are
\begin{eqnarray}
{\lambda}_{00} &=&
\frac{\epsilon(2-\epsilon)}{1+\frac{4}{5}(1-\epsilon)}\\
{\omega}_{000} &=&
\left[
\frac{3\epsilon(2-\epsilon)+2\epsilon^2(3-2\epsilon)
  +\frac{24}{5}\epsilon(1-\epsilon)^2}
{1+\frac{20}{7}(1-\epsilon)^3+\frac{12}{5}(1-\epsilon)^2\epsilon}
 - 3{\lambda}_{00}
\right]\frac{{\lambda}_{00}^{-\frac{3}{2}}}{2}.
\end{eqnarray}
These are shown as functions of $\epsilon$ 
in Figs. \ref{fig:chap} and \ref{fig:wfp}.
The chaoticity for the normalized correlations of differently charged pions
vanishes, because the generic part exhibits no HBT effect and the DCC part is
taken to be coherent.

It should be noted that the distribution function of the fractional
neutral-pion number in the partially coherent DCC model is different
from Eq. (\ref{eq:dist}).
The fractional neutron-pion number in the partially coherent DCC model is
\begin{eqnarray}
f' &=& \frac{W_f^0(p_1)}{W_f^+(p_1)+W_f^-(p_1)+W_f^0(p_1)}\nonumber\\
&=& \frac{f\cdot f_{11}+F_{11}}{f_{11}+3F_{11}} \nonumber\\
&=& f(1-\epsilon)+\frac{1}{3}\epsilon,
\end{eqnarray}
where each kind of pion emitted from the generic source is assumed to have
the same multiplicity, or $F_{11}^0=F_{11}^+=F_{11}^-$.
$\epsilon$ is that of Eq. (\ref{eq:ratio}).
The new probability distribution function for $f'$, $P'(f')$
is obtained by the use of $P(f)df=P'(f')df'$ as
\begin{equation}
P'(f') =
\left\{
\begin{array}{lc}
 \displaystyle
 \frac{1}{2\sqrt{(f'-\frac{1}{3}\epsilon)(1-\epsilon)}}
&
(\frac{1}{3}\epsilon < f' < 1-\frac{2}{3}\epsilon)
\\
0 & \mbox{(otherwise)}
\end{array}\right. .
\end{equation}
This distribution function becomes Eq. (\ref{eq:dist}) when $\epsilon = 0$, 
or one DCC source with no chaotic source.
When $\epsilon = 1$, or no DCC source,
$P'(f')$ is meaningful only at $f'=\frac{1}{3}$.
The charge fluctuation caused by DCC is reduced as the generic source increases.

\subsection{Multiple DCC Domains (M-DCC)}
\label{sec:sdcc}

In actual experiments, we expect that DCC would appear not simply in a 
possibly large, unified area but in several, separated domains.
In order to describe such a domain structure, we use multiple coherent sources 
whose number obeys the Poisson distribution.

In an analogy to the description of a generic source in \cite{gkw},
we write the source current for $N$ DCC domains as
\begin{equation}
J_i(p) = \sum_{n=1}^Nj(p)e^{ip\cdot X_n-i\theta_n}n_{n,i},
\end{equation}
where $n_{n,i}$ is a unit-vector in isospin space,
describing the direction of condensate in the $n$-th domain.
The position of the $n$-th domain is $X_n$,
distributed according to the function, $\rho(X_n)$.
The normalization of $\rho(x)$ is $\int\rho(x)dx =1$.
$\theta_n$ is a random number uniformly distributed between 0 and $2\pi$,
so as to satisfy the chaotic property among the different domains.
$N$ obeys the Poisson distribution, but the zero-domain event, or $N=0$,
should be excluded from it.
Instead of the usual Poisson distribution, the appropriately renormalized
Poisson distribution is then
\BE
\label{eq:poisson}
{\gamma}_N^{(renorm)}=\frac{\alpha^N}{N!}\frac{1}{1-e^\alpha}
\quad \mbox{for } N = 1 \sim \infty.
\EE
In this case, the average of the generating functional, Eq. (\ref{eq:gf}),
 is
\BE
\langle\cdots\rangle_J
=
\sum_{N=1}^\infty {\gamma}_N^{(renorm)}\int \left(\prod_{n=1}^N
\frac{d^3{\mathbf{n}}_{n}}{4\pi}
\delta(|{\mathbf{n}}_{n}|-1) \right)
\int \left(\prod_{n=1}^N dX_N \rho(X_N) \right)
\int_0^{2\pi}\left(\prod_{n=1}^N \frac{d\theta_n}{2\pi}\right)\cdots.
\EE

The pion spectra are obtained as 
\begin{mathletters}
\BEA
W^+_1(p_1) &=& \frac{\alpha}{3}\frac{|j(p_1)|^2}{1-e^{-\alpha}},\\
W^{++}_2(p_1,p_2) &=& \frac{|j(p_1)|^2|j(p_2)|^2}{1-e^{-\alpha}}
\left[
\frac{2}{15}\alpha+\frac{1}{9}\alpha^2+\frac{1}{9}\alpha^2
|\rho_{12}|^2 
\right]
\\
W^{+++}_3(p_1,p_2,p_3) &=& \frac{|j(p_1)|^2|j(p_2)|^2|j(p_3)|^2}
        {1-e^{-\alpha}} \nonumber,\\
&
\times
&
\left[
\frac{2}{35}\alpha+\frac{2}{15}\alpha^2+\frac{1}{27}\alpha^3
\right. \nonumber \\ && \left.
+\left(\frac{4}{45}\alpha^2+\frac{1}{27}\alpha^3\right)
\sum_{(a,b)}|\rho_{ab}|^2+\frac{1}{27}\alpha^3\cdot
2 {\mathrm{Re}}(\rho_{12}\rho_{23}\rho_{31})
\right],
\EEA
\end{mathletters}
where
\BE
 \rho_{ij}=\int d^4x \rho(x)e^{-i(p_i-p_j)\cdot x}.
\label{eq:rho}
\EE
Using the above equations, we derive 
the chaoticity and weight factor for normalized correlations as
\BEA
{\lambda}_{++} &=& \frac{\alpha}{\alpha+\frac{6}{5}},\\
{\omega}_{+++}&=& \frac{\alpha^2+\frac{6}{5}\alpha+\frac{351}{175}}
{\alpha^2+\frac{18}{5}\alpha+\frac{54}{35}}\sqrt{\frac{\alpha+\frac{6}{5}}
{\alpha}}.
\EEA
If the domains are not DCC sources but coherent generic sources,
the chaoticity and weight factor become\cite{ns2}
\BEA
\lambda_{generic} &=& \frac{\alpha}{\alpha+1} ,\\
\omega_{generic} &=& \frac{1}{2}\frac{2\alpha^2+2\alpha+3}{\alpha^2+3\alpha+1}
  \sqrt{\frac{\alpha+1}{\alpha}},
\EEA
respectively.
In Figs. \ref{fig:chas} and \ref{fig:wfs}, we show the chaoticity and
weight factor of the multiple DCC model (M-DCC)  
as functions of the mean number of the DCC domains, $\langle N \rangle=
\alpha/(1-\exp(-\alpha))$.  For comparison, we also show in the figure 
the chaoticity and weight factor of the multiple coherent generic
model (M-G)\cite{gkw} that we examined in our previous work\cite{ns2}. 

In this model, the correlations of differently charged pions has no HBT effect.
For example, the two-pion correlation function of a paired positive pion
and negative pion is 
\BEA
W_2^{+-}(p_1,p_2) &=& \frac{|j(p_1)|^2|j(p_2)|^2}{1-e^{-\alpha}}
\left[\frac{2}{15}\alpha+\frac{1}{9}\alpha^2\right]. 
\EEA
$W_2^{+-}(p_1,p_2)$ is independent of $\rho_{12}$.
Note that $n_{n,+}^2$ appears during the calculation, but it is reduced
to no contribution after averaging over the azimuthal angle by taking account
of the isotropic distribution of condensate:
\BE
\int_0^1 P(f)df \int_0^{2\pi}\frac{d\phi}{2\pi} n_{+}^2 
 = \int \frac{1-f}{2}P(f)df \int \exp(2i\phi)\frac{d\phi}{2\pi}
= 0 ,
\EE
where $n_1=\sqrt{1-f}\cos\phi$, $n_2=\sqrt{1-f}\sin\phi$.
The HBT effect thus vanishes for a paired positive pion and negative
pion by this averaging.  Similarly, the correlation of a neutral pion and
a charged pion shows no HBT effect.

The chaoticity and weight factor of neutral pions are
\BEA
\lambda_{00} &=& \frac{\alpha}{\alpha+\frac{9}{5}} \\
\omega_{000} &=& \frac{1}{2}
  \frac{2\alpha^2+\frac{18}{5}\alpha+\frac{1377}{175}}
      {\alpha^2+\frac{27}{5}\alpha+\frac{27}{7}}
  \sqrt{\frac{\alpha+\frac{9}{5}}{\alpha}}.
\EEA
These are illustrated as functions of the mean number of domains 
in Figs. \ref{fig:chas} and \ref{fig:wfs}.

As in the case of the partially coherent DCC of the previous subsection,
the fractional pion number is also modified in the case of multiple DCC domains.
We can write the probability distribution of $f$ for $N$ DCC domains as
\BE
P_N(f) = \int_0^1\delta\left(f-\frac{1}{N}\sum_{n=1}^Nf_n\right)
\frac{1}{2\sqrt{f_1}}\cdots\frac{1}{2\sqrt{f_N}}
df_1df_2\cdots df_N,
\EE
where $f_n$ is the ratio of the neutral pion number and the total pion number 
in the $n$-th domain.
The pion spectra emitted from each domain is assumed to be the same.
The observed distribution is obtained by averaging the above
probability distribution with the Poisson distribution, Eq. (\ref{eq:poisson}),
\BE
P_{\alpha}(f) = \sum_{N=1}\frac{\alpha^N}{N!}\frac{1}{e^\alpha-1}P_N(f).
\EE
Figure \ref{fig:pfs} illustrates $P_{\alpha}(f)$ for the mean
number of the domains, $\langle N \rangle = 1, 2, 4$, and 20.
$\langle N \rangle=1$ corresponds to Eq. (\ref{eq:dist}).
Figure \ref{fig:pfs} is similar to a figure in {}\cite{gavin}, 
where no detail of the derivation is given.
The distribution indeed becomes sharper as  $\langle N \rangle$ 
increases\cite{gavin,rajagopal}.

\subsection{Multiple DCC Domains with a Chaotic Generic Source (M-DCC/G)}

We now consider the case of multiple DCC domains with one
chaotic source emitting generic pions.
This model corresponds to the combination of the two models
in the previous subsections.
Note that the Poisson distribution used in the current model is the usual one,
instead of Eq. (\ref{eq:poisson}):
\BE
{\gamma}^{(usual)}_N=\frac{\alpha^N}{N!}e^{-\alpha}.
\EE

The source current is
\BE
J_i(p) = \sum_{n=1}^N j(p) e^{ip\cdot X_n -i\theta_n}n_{n,i}+J^i_g(p),
\EE
where the variables are defined in the same way as in the previous
subsections.
For convenience, we introduce $S^i_{ab}$ and $T_{ab}$:
\BEA
T_{ab}&=& \frac{j^*(p_a)j(p_b)}{2(2\pi)^3\sqrt{p_a^0p_b^0}}\rho_{ab} ,\\
S^i_{ab}&=& \langle J_g^{i*}(p_a)J_g^i(p_b) \rangle_{J_g^i},
\EEA
where $\rho_{ij}$ is defined in Eq. (\ref{eq:rho}) and
$S^i_{ab}$ is equivalent to $F_{ab}^i\exp(i\Phi_{ab}^i)$ in Eq. (\ref{eq:fij}).
Hereafter, when we do not denote the superscript $i$ explicitly,
we imply that $+$ is excluded.

The one-pion spectrum and two- and three-pion correlations are
\begin{mathletters}
\BEA
W_1^+(p_1) &=& \frac{\alpha}{3}T_{11}+S_{11} ,\\
C_2^{++}(p_1,p_2) &=& 1
+\frac{\left|\frac{\alpha}{3}T_{12}+S_{12}\right|^2}
      {\left[1+\frac{6}{5}\frac{(1-\epsilon)^2}{\alpha}\right]
        W_1^+(p_1)W_1^+(p_2)} ,\\
C_3^{+++}(p_1,p_2,p_3)&=& 1+ \frac{1}
   {1+\frac{54}{35}\frac{(1-\epsilon)^3}{\alpha^2}
          +\frac{18}{5}\frac{(1-\epsilon)^2}{\alpha}}
\nonumber \\ && \times
 \left\{\sum_{(a,b)}
   \frac{\left|\frac{\alpha}{3}T_{ab}+S_{ab}\right|^2
      +\frac{4}{15}\alpha(1-\epsilon)\left|T_{ab}\right|^2
          +\frac{2}{5}(1-\epsilon)\cdot 2{\mathrm{Re}}(T_{ab}S_{ba})}
        {W^+_1(p_a)W^+_1(p_b)}
\right. \nonumber \\ && \left.
   +\frac{2{\mathrm{Re}}\left[
         \left(\frac{\alpha}{3}T_{12}+S_{12}\right)
         \left(\frac{\alpha}{3}T_{23}+S_{23}\right)
         \left(\frac{\alpha}{3}T_{31}+S_{31}\right)
         \right]}
        {W^+_1(p_1)W^+_1(p_2)W^+_1(p_2)}\right\},
\EEA
\end{mathletters}
where the ratio of the generic pion number and the total pion number,
$\epsilon$, is
\BE
 \epsilon = \frac{S_{11}}{\frac{\alpha}{3}T_{11}+S_{11}}.
\EE
We assume that $\epsilon$ is independent of $p$.
The chaoticity and weight factor are obtained as
\BEA
\lambda_{++} &=& \frac{\alpha}{\alpha+\frac{6}{5}(1-\epsilon)^2} \\
\omega_{+++} &=& \frac{\alpha^2+\frac{6}{5}\alpha(1-\epsilon)^2
    +\frac{27}{175}(1-\epsilon)^3(13-28\epsilon)}
    {\alpha^2+\frac{18}{5}\alpha(1-\epsilon)^2+\frac{54}{35}(1-\epsilon)^3}
 \sqrt{\frac{\alpha+\frac{6}{5}(1-\epsilon)^2}{\alpha}}.
\EEA
In Figs. \ref{fig:dchal} and \ref{fig:dchaw},
the chaoticity and weight factor are shown
as functions of the mean number of DCC domains,
$\langle N \rangle = \alpha$, for various values of $\epsilon$.
     
The chaoticity and weight factor of neutral pions are obtained as
\BEA
\lambda_{00} &=& \frac{\alpha}{\alpha+\frac{9}{5}(1-\epsilon)^2} \\
\omega_{000} &=& \frac{1}{2}
  \frac{2\alpha^2+\frac{18}{5}\alpha(1-\epsilon)^2
         +\frac{81}{175}(1-\epsilon)^3(17-42\epsilon)}
      {\alpha^2+\frac{27}{5}\alpha(1-\epsilon)^2
        +\frac{27}{7}(1-\epsilon)^3}
  \sqrt{\frac{\alpha+\frac{9}{5}(1-\epsilon)^3}{\alpha}}.
\EEA
where the spectra of neutral generic pions is 
the same as that of the positive ones,
or $S^+_{11}=S^0_{11}$.
The correlation functions of nonidentical pions
have no HBT effect, as in the two previous models.

If $N$ DCC domains are produced,
the fractional neutral-pion number is actually 
\BEA
f&=&\frac{\sum_{n}^Nf_nT_{11}+S^0_{11}}
   {NT_{11}+S^0_{11}+S^+_{11}+S^-_{11}} \nonumber \\
&=& \frac{\sum_{n}^N3f_n(1-\epsilon)+\epsilon\alpha}
   {3N(1-\epsilon)+3\epsilon\alpha}.
\EEA
As a consequence, the probability distribution of $f$ is 
\BE
\label{eq:disae}
 P_{\alpha,\epsilon}(f) = \sum_{N=0}^\infty \frac{\alpha^N}{N!}e^{-\alpha}
  \int \delta\left( f - \frac{\sum_{n}^N3f_n(1-\epsilon)+\epsilon\alpha}
   {3N(1-\epsilon)+3\epsilon\alpha}\right)
   \prod_{n=1}^N\frac{df_n}{2\sqrt{f_n}}.
\EE

\section{Discussions and Summary}
\label{sec5}

\begin{figure}
\begin{center}
\epsfig{file=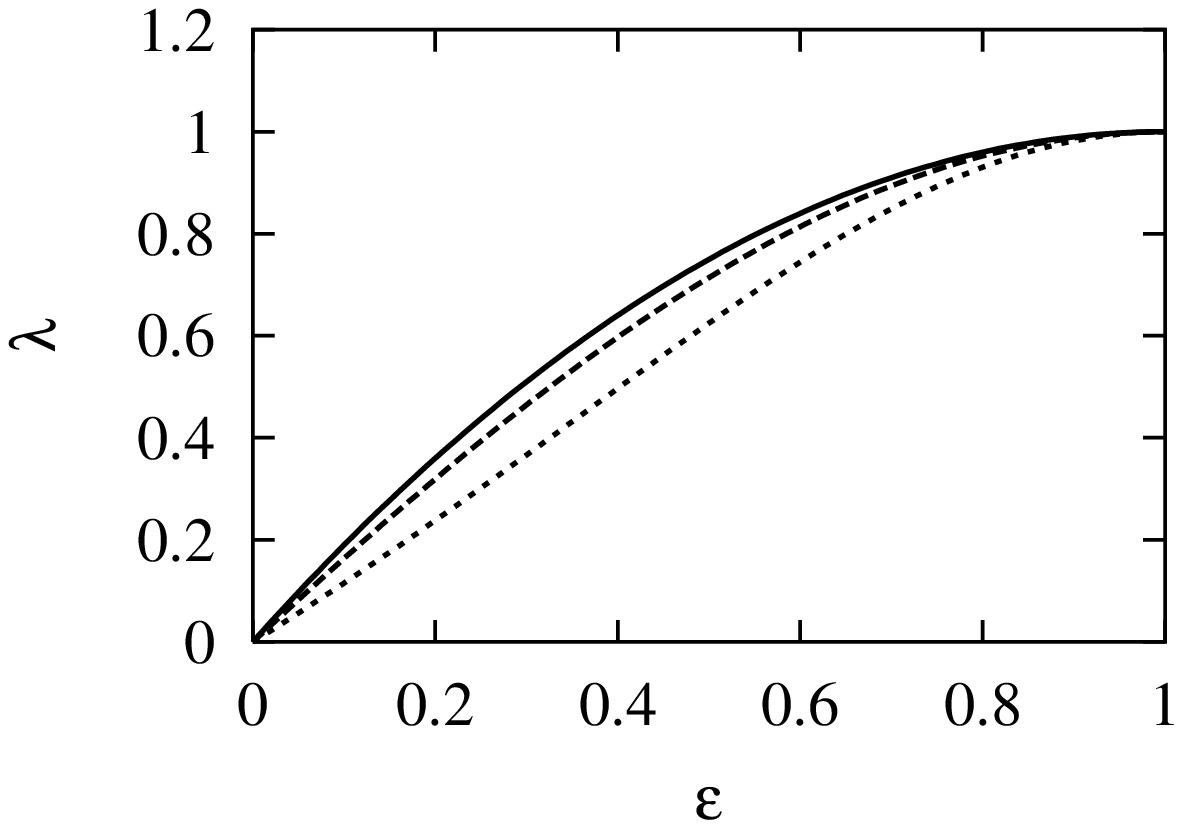 }%
\caption{Chaoticity, $\lambda$, as a function of the ratio of the generic pion 
number and the total pion number, $\epsilon$. 
The solid curve represents the partially coherent generic model (PC-G). 
The dashed and dotted curves represent the partially coherent DCC model 
(PC-DCC) for the positive pions and for the neutral pions, respectively.}
\label{fig:chap}
\end{center}
\end{figure}

\begin{figure}
\begin{center}
\epsfig{file=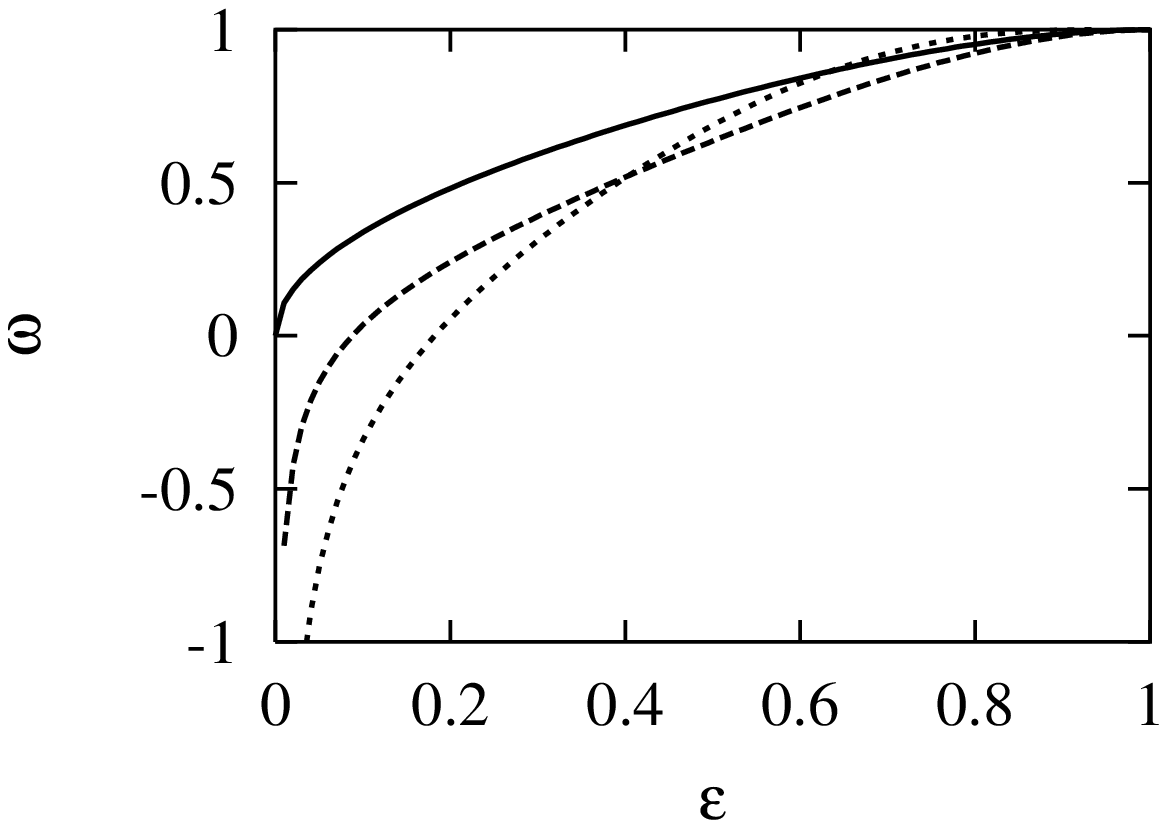}%
\caption{Weight factor, $\omega$, as a function of the ratio of 
the generic pion number and the total pion number, $\epsilon$. 
The solid curve represents the partially coherent generic model (PC-G). 
The dashed and dotted curves represent the partially coherent DCC model 
(PC-DCC) for the positive pions and for the neutral pions, respectively.}
\label{fig:wfp}
\end{center}
\end{figure}

\begin{figure}
\begin{center}
\epsfig{file=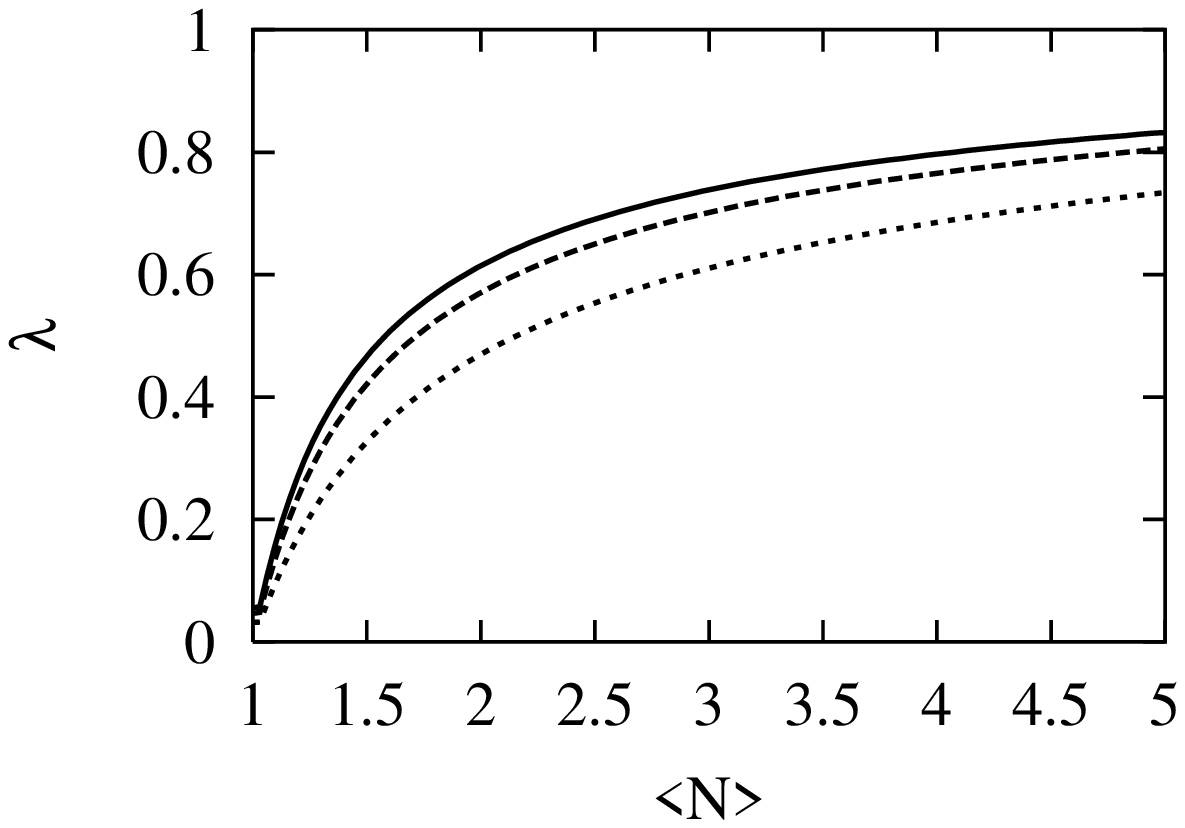}%
\caption{Chaoticity, $\lambda$, as a function of the mean number of domains, 
$\langle N \rangle$.  
The solid curve represents the multiple coherent generic sources (M-G). 
The dashed and dotted curves represent the multiple DCC domain model (M-DCC) 
for the positive pions and for the neutral pions, respectively.}
\label{fig:chas}
\end{center}
\end{figure}

\begin{figure}
\begin{center}
\epsfig{file=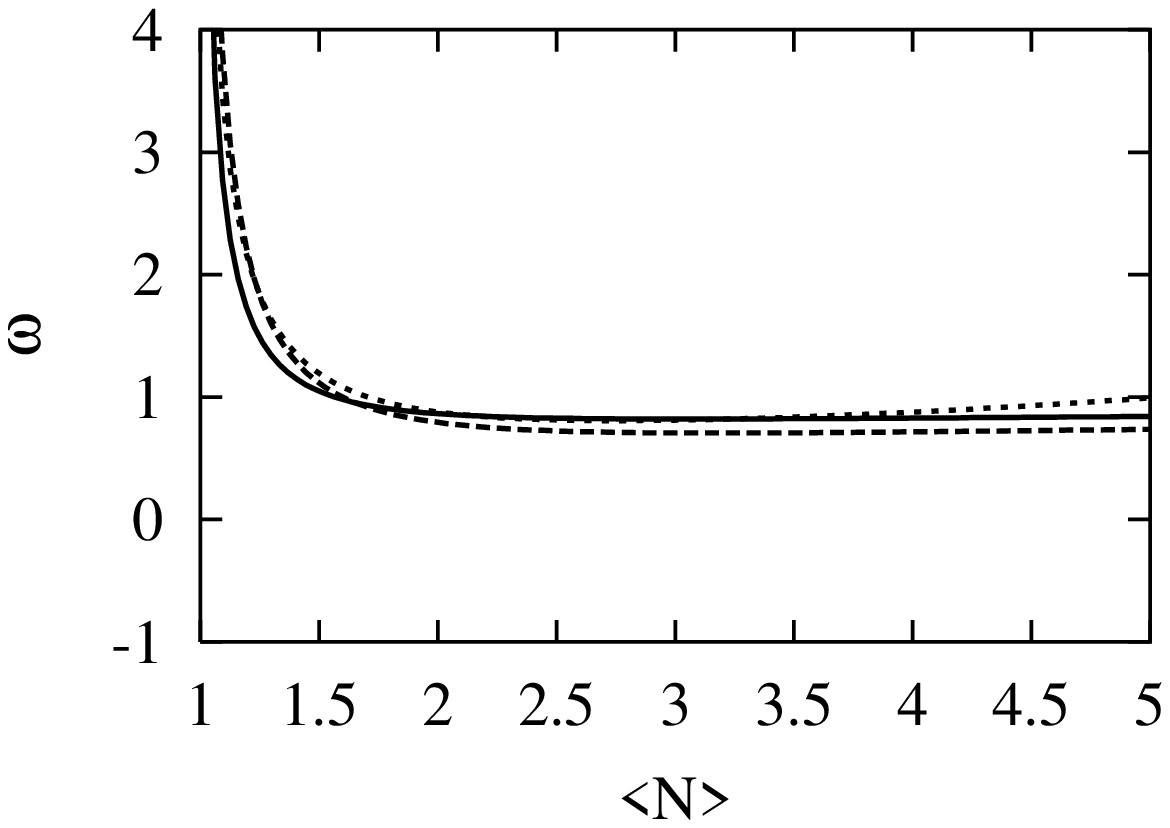}%
\caption{Weight factor, $\omega$, as a function of the mean number of domains, 
$\langle N \rangle$.  
The solid curve represents the multiple coherent generic sources (M-G). 
The dashed and dotted curves represent the multiple DCC domain model (M-DCC) 
for the positive pions and for the neutral pions, respectively.}
\label{fig:wfs}
\end{center}
\end{figure}

\begin{figure}
\begin{center}
\epsfig{file=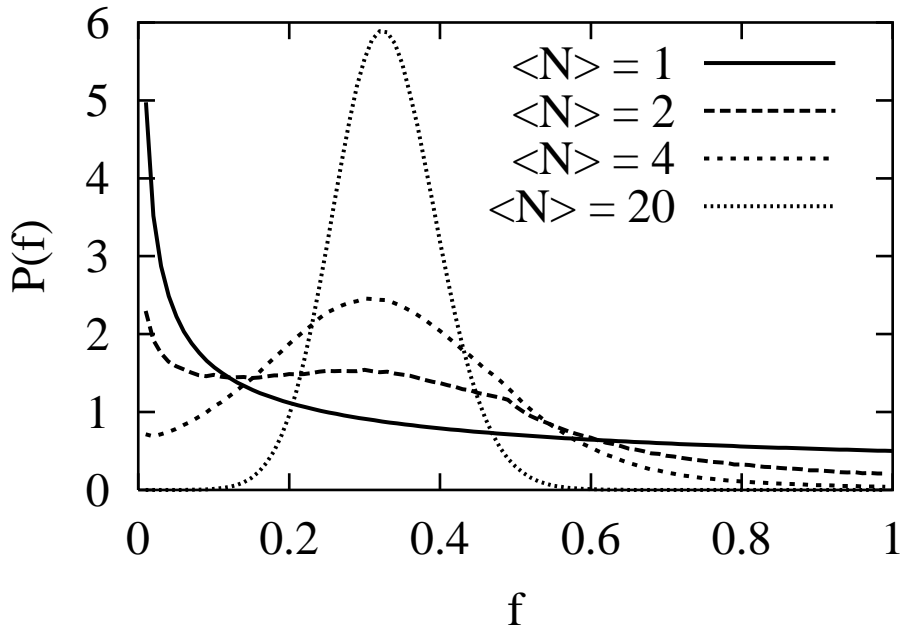}%
\caption{Probability distributions in the multiple DCC domain model, 
$P_{\alpha} (f)$, as a function of the fractional neutral-pion number, $f$,  
the mean domain number, $\langle N\rangle = 1,2,4,20$, 
(corresponding to $\alpha = 0, 1.59, 3.92, 20.0$, respectively).}
\label{fig:pfs}
\end{center}
\end{figure}

\begin{figure}
\begin{center}
\epsfig{file=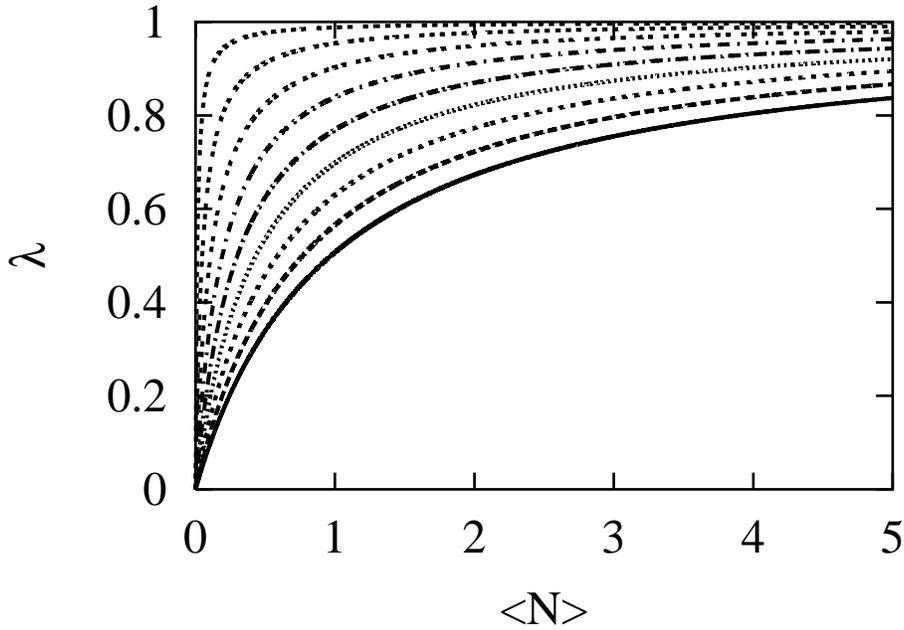}
\caption{Chaoticity, $\lambda$, of the positive pions 
for the model of multiple DCC domains with one chaotic generic source,
as a function of the mean number of DCC domains, $\langle N\rangle$.
The curves are shown for the ratio of 
the generic pion number and the total pion number, $\epsilon=0.1-0.9$,
with step 0.1 from down to up.
}
\label{fig:dchal}
\end{center}
\end{figure}

\begin{figure}
\begin{center}
\epsfig{file=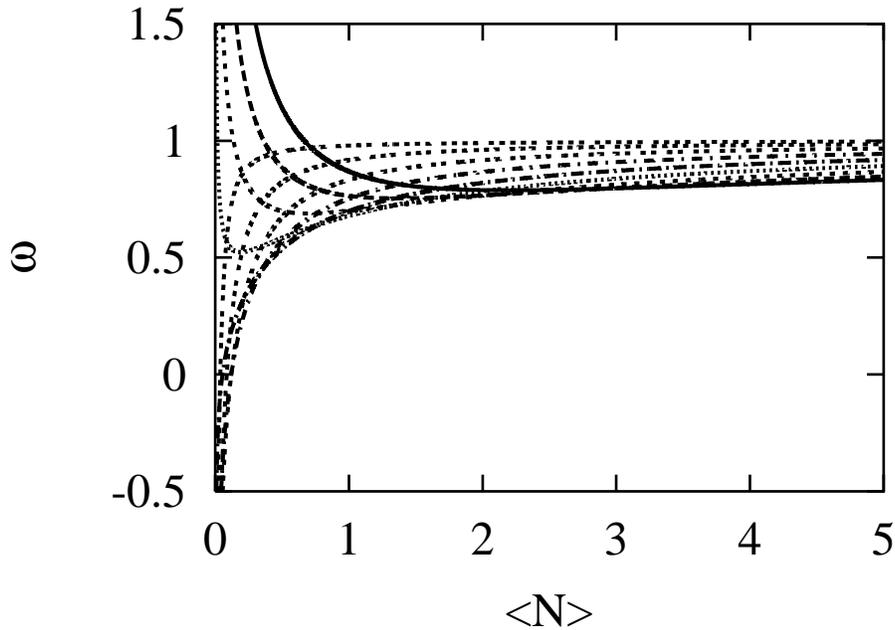}
\caption{Weight factor, $\omega$, of the positive pions for the model of 
multiple DCC domains with one chaotic generic source (M-DCC/G), 
as a function of the mean number of DCC domains, $\langle N\rangle$, 
varying the ratio of the generic pion number and the total pion number, 
$\epsilon$, from 0.1 to 0.9 with step 0.1, from down to up
at $\langle N\rangle=5$.}
\label{fig:dchaw}
\end{center}
\end{figure}

As we have seen in Sec. III, the pions emitted from a single coherent source 
of DCC do not yield pion interferometry useful for identifying 
a signature of DCC.    
Also, when the pions are emitted from a source in which a single coherent DCC 
source and a chaotic source are combined, 
this situation remains practically the same.   
Figures \ref{fig:chap} and \ref{fig:chas} show that the partially coherent 
DCC source (PC-DCC) and the partially coherent generic source (PC-G) 
each yield a chaoticity and weight factor similar to that of the other. 
The difference between the two is numerically too close for interferometry 
to serve as a means identifying a DCC signature. 
If pion interferometry for the neutral pions should become feasible, 
combined data of charged and neutral-pion interferometry would serve as 
a means.  Note that the pion interferometry of different charges 
exhibits no HBT effect and is not useful.

\begin{figure}
\begin{center}
\epsfig{file=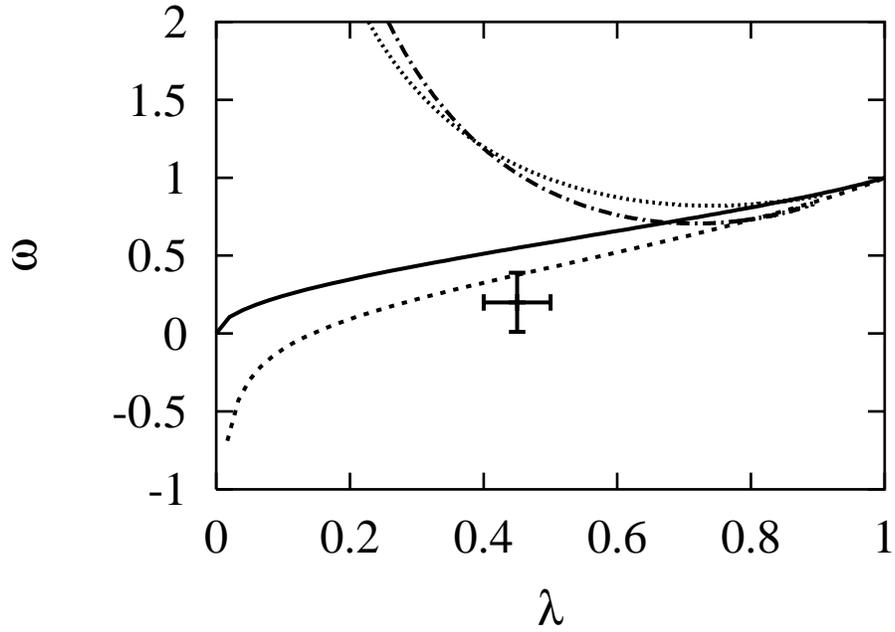}%
\caption{Weight factor as a function of chaoticity for the positive pions.
The solid and dashed curves represent the partially coherent generic model 
(PC-G) and the partially coherent DCC model (PC-DCC).
The dotted  and dot-dashed curves represent the multiple coherent generic model 
(M-G) and the multiple DCC domain model (M-DCC), respectively. 
The data point is that from the CERN NA44 [7]. 
}
\label{fig:cvswf}
\end{center}
\end{figure}

Figure \ref{fig:cvswf}
 illustrates the weight factor as a function of the chaoticity 
in the models examined in this work, except for the model of multiple 
DCC domains with one generic source (M-DCC/G). 
For comparison, we also show in this figure the recent experimental data 
from the CERN NA44 Collaboration\cite{na44}, the chaoticity and weight factor 
being $0.4-0.5$ and $0.20\pm0.19$, respectively.
(We have combined the systematic and statistical uncertainties 
quadratically.) 
The partially coherent DCC model (PC-DCC) is closest to the data point, 
while  the partially coherent generic model (PC-G) is next. 
The ratio of the generic pion number and the total pion number, $\epsilon$, 
that yields the closest PC-DCC is found from Fig. 2 
to be about 0.30.  
That is, about 70\% of pions are emitted from the DCC domains.  This implies 
that we have a large charge fluctuation of the pions emitted.  
Such a large charge fluctuation does not seem to have been observed 
in the experiments\cite{wa98}, however.   Note that the essential aspect 
of the model that yields the results close to the data is the inclusion 
of a chaotic generic source.  In Fig. \ref{fig:cvswf},
 we see the model of multiple DCC 
domains (M-DCC) with no chaotic generic source yielding the weight factor  
far away from the data point.

\begin{figure}
\begin{center}
\epsfig{file=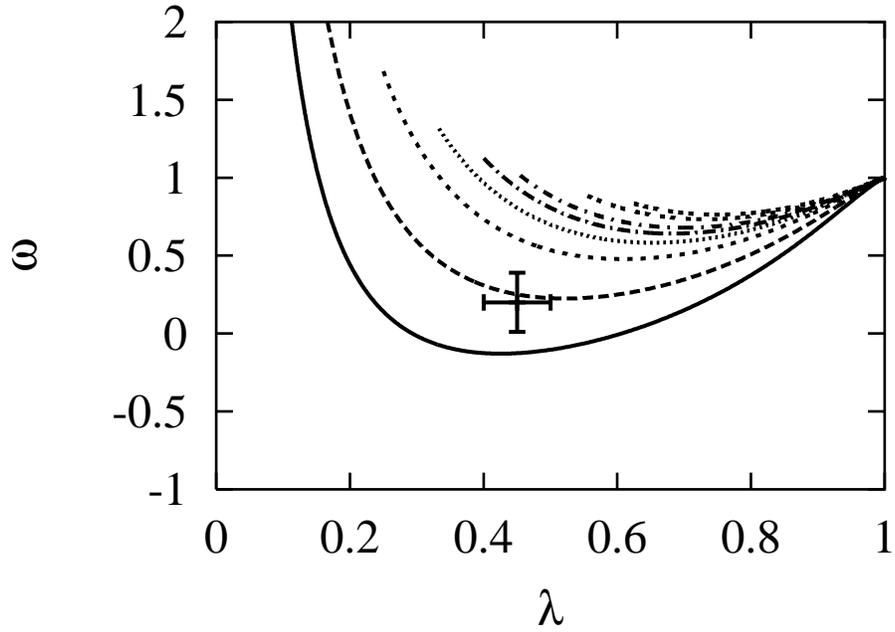}
\caption{Weight factor, $\omega$, as a function of chaoticity, $\lambda$, 
for the positive pions, for the model of multiple DCC domains with one 
generic chaotic source (M-DCC/G), varying 
$\epsilon$ from 0 to 1.
The lines from down to up correspond to the mean number of domains,
$\langle N \rangle = 0.1$, 0.2, 0.4, 0.6, 0.8, 1.0, 1.5 and 2.0, respectively.
The data is from the NA44 experiment [7].
}
\label{fig:dchalw}
\end{center}
\end{figure}

Figure \ref{fig:dchalw} shows that the multiple DCC domain model (M-DCC/G) 
successfully yields the chaoticity and weight factor in agreement with the 
data.  The parameter values for the best fit are found to be 
$\alpha = \langle N \rangle = 0.18$ and $\epsilon=0.57$ from Fig. 7.
This implies that the mean number of the DCC domains
is 0.18 and that the ratio of the generic pion number and the total pion 
number is 0.57.  For these parameter values, 
the charge distribution, Eq. (\ref{eq:disae}), is 
\BE
P_{\alpha,\epsilon}(f) \approx 0.83 P_0(f) + 0.15 P_1(f).
\label{eq:distex}
\EE
Here, $P_0(f)$ is equal to $\delta(f - 1/3)$ in our simple model, 
while $P_1(f)$ is approximately the inverse square root of $f$, 
as in Eq. (\ref{eq:dist}).  Since $P_1(f)$ is suppressed by the factor of 
0.15, $P_{\alpha,\epsilon}(f)$ of Eq. (\ref{eq:distex}) is a distribution 
dominated by a sharp peak with a slow-varying $1/\sqrt{f}$-like background. 
In practice, the sharp peak can be replaced by a smoother function, 
such as a binomial distribution peaking at $f=1/3$.

We have thus shown that the NA44 data can be explained by the model of 
multiple DCC with a chaotic generic source.  We have not demonstrated, 
however, that the NA44 data prove the appearance of DCC.  Unfortunately, 
if we limit ourselves to pion interferometry, we require 
interferometry of the neutral pions in order to strengthen the case.   
For the above parameter values, the chaoticity and weight factor for the 
neutral pions are $\lambda_{00}=0.35$ and $\omega_{000} = -0.12$, respectively.
If interferometry for the neutral pions should be carried out, these 
values should signal the observation of DCC.  Note that pion 
interferometry of differently charged pions does not serve for identifying 
DCC because the generic part exhibits no HBT effect and the multiple DCC part 
also has no HBT effect due to the azimuthal average in isospin space
(as discussed in the case of the partially coherent 
DCC model in Subsec. \ref{sec:pcdcc} and in the case of multiple DCC
domain model in Subsec. \ref{sec:sdcc}).

In summary,
we have investigated two- and three-pion correlations when DCC occurs.
A chaoticity and weight factor are used as measures
of two- and three-pion correlations and
the various models for DCC have been investigated.
The chaoticities and weight factors  for DCC are different from 
the generic case. 
The existing experimental data are in agreement with
the model of multiple DCC domains with one generic source.
We suggest that the chaoticity and weight factor for neutral pions
together with those for the charged pions should enable the identification
occurrences of DCC.

\acknowledgements
We acknowledge informative and stimulating discussions with T. Humanic, 
especially regarding the NA44 experiment.
This research is supported 
by the U.S.~Department of Energy under grant DE-FG03-87ER40347 at CSUN, 
and the U.S.~National Science
Foundation under grants PHY88-17296 and PHY90-13248 at Caltech.

\newpage

\newpage

\end{document}